# Discovery and Engineering of Low Work Function Perovskite Materials

Tianyu Ma[1], Ryan Jacobs[1], John Booske[2], and Dane Morgan[1]

[1]Department of Materials Science and Engineering, University of Wisconsin-Madison, Madison, WI, 53706, USA

[2]Department of Electrical and Computer Engineering, University of Wisconsin-Madison, Madison, WI, 53706, USA

Corresponding author E-mail: ddmorgan@wisc.edu

**Abstract**

Materials with low work functions are critical for an array of applications requiring the facile removal or efficient transport of electrons through a device. Perovskite oxides are a promising class of materials for finding low work functions, and here we target applications in thermionic and field electron emission. Perovskites have highly malleable compositions which enable tunable work function values over a wide range, robust stability at high temperatures, and high electronic conductivities. In this work, we screened over 2900 perovskite oxides in search of stable, conductive, low-work-function materials using Density Functional Theory (DFT) methods. Our work provides insight into the materials chemistry governing the work function value of a perovskite, where materials with barely filled *d* bands possess the lowest work functions. Our screening has resulted in a total of seven promising compounds, such as $BaMoO_3$ and $SrNb_{0.75}Co_{0.25}O_3$ with work functions of 1.1 eV and 1.5 eV, respectively. These promising materials and others presented in this study may find use as low work function electron emitters in high power vacuum electronic and thermionic energy conversion devices. Moreover, the database of calculated work functions and materials chemistry trends governing the value of the work function may aid in the engineering of perovskite heterojunction devices.

## 1 Introduction

The work function is a fundamental surface electronic property, defined as the energy required to move an electron from the bulk Fermi level to the vacuum level outside the material.

The work function is a crucial parameter for materials used in a wide array of vacuum and solid-state electronic devices, including vacuum electronics, thermionic energy converters, oxide electronic heterostructures, solar photovoltaic cells, memristors and catalysts. In this work we target application of perovskites for thermionic and potentially field electron emission cathodes. A low work function material is critical in emission cathodes for the generation of the high emitted current density[1], [2] needed for the efficient use of thermionic, photo, and field emitters widely employed in high power, high frequency vacuum electronic device (VED) applications such as traveling wave tubes (TWTs),[3] klystrons,[4] gyrotrons,[5] and magnetrons.[6] These devices are crucial components in diverse applications ranging from communications, industrial food production, national defense, deep space exploration, and scientific research.[7] The work function is also the crucial parameter for electron emissive layers employed in conventional and photon-enhanced thermionic energy conversion devices, where both high and low work function materials are desirable in order to maximize the open circuit voltage of the thermionic converter, thus enabling high output power for a given current density.[8]–[10]

Impregnated W cathodes have been used as a thermionic emission material in commercial VEDs for decades. In its simplest form, the impregnated W cathode consists of a porous tungsten body impregnated with electron emission promoting materials such as a $CaO-Al_2O_3-BaO$ mixture (the so-called "B-type" cathode). During activation, free Ba atoms diffuse to the W surface and form a Ba-O monolayer[11], which acts as a surface dipole[12] and lowers the effective work function of the cathode from 4.5 eV (the average value of bare polycrystalline W) to about 2 eV.[13] The B-type cathode has been further engineered to include an Os/Ru thin film coating, resulting in a further reduced work function of about 1.8 eV (the so-called "M-type" cathode).[14] The newest generation of impregnated W cathode is the scandate cathode (W + BaO + $Sc_2O_3$). These cathodes have gained much research interest in the past 25 years due to their extremely low work function of about 1.1 eV and ability to realize ultra-high emitted current densities in excess of 100 $A/cm^2$.[15]–[19] Despite the commercial use of B- and M-type cathodes and decades of research on scandate cathodes, impregnated cathodes of all types have important shortcomings. Namely, these cathodes suffer from Ba depletion[20], [21] and surface degradation[14] as a result of the continual evaporation of Ba from the emitting surface during high temperature operation. This off-gassing of the work-function-lowering Ba species limits the cathode lifetime. In addition, impregnated cathodes can experience non-uniform Ba coverage, leading to device efficiency limits[22] and engineering challenges.[23] For the specific case of scandate cathodes, the previous shortcomings all apply, but scandate cathodes tend to require extra activation and have issues of manufacturing and performance repeatability.[19], [24]–[26] These above-mentioned issues all stem from the fact that impregnated cathodes realize their low work function through the formation of a volatile surface dipole layer believed to be comprised mainly of Ba and O. While much study has been devoted to understanding the surface chemistry, atomic-scale structures and resultant work functions of the surface dipole species for B-type, M-type and scandate cathodes,[12], [27]–[31] the natural materials design question becomes: is there a different material one could use as an electron emitter that can reach low work function without extra surface layer, and therefore not be prone to the critical issue of limited cathode lifetime? The discovery and design of such a stable, low work function, ultra-long-lifetime electron emission material may ultimately enable the deployment of high-power, high frequency VEDs and related devices that can operate at lower temperatures, realize high emitted current densities, and reduce operational and device replacement costs.

To address the above materials design question, in this study we turn our attention to perovskite oxide materials ($A_{1-x}A_x'B_{1-y}B_y'O_3$ stoichiometry). In general, perovskites are a broad class of materials known for their high degree of compositional flexibility, where their structure can support upwards of 90% of the elements in the periodic table (within some solubility limit) without losing its three-dimensional interconnected network of octahedra.[32] This high degree of compositional flexibility enables perovskites to have a wide range of tunable properties, including tunable work function values. The recent studies of Jacobs *et al.*[33] and Zhong *et al.*[34] both demonstrated that the work function of perovskite materials can vary on the scale of several eV, and depends on the bulk material composition, the surface orientation (e.g. (001) versus (011)), and the surface termination (e.g. AO- versus $BO_2$-terminated (001) surface). These studies were the first to characterize the work functions of a representative set of perovskite oxides using Density Functional Theory (DFT) methods for applications in electron emission and heterojunction engineering, and suggest that perovskites with low work functions may be realized by virtue of the polar (001) oriented surfaces of the perovskite structure.[33], [34] Jacobs *et al.* also identified an approximate linear correlation between (001) surface work function and the bulk O *p*-band center.[33] A recent study from Xiong *et al.*[35] used data-driven approach to analyze the electronic-structure factors controlling the work function of perovskites, and corroborated that the O *p*-band center is an effective work function descriptor. The O *p*-band center is the centroid of the oxygen-projected density of states (see **Section 4** for more details), and can be calculated from a single DFT bulk structure relaxation run. This established correlation of surface work function with bulk O *p*-band center enables fast screening of perovskite work functions at a rate that is approximately two orders of magnitude faster than conducting computationally expensive Heyd-Scuseria-Ernzerhof (HSE)[36] hybrid functional calculations of perovskite surface slabs. Jacobs *et al.* specifically identified $SrVO_3$ as a promising cathode material with a calculated (001) SrO terminated work function of about 1.9 eV and predicted that Ba dopant impurities in $SrVO_3$ should segregated onto the (001) surface, creating a BaO-terminated (001) surface with a very low work function of 1.1 eV[33]. Meanwhile, $SrVO_3$ has been experimentally shown to be a good electronic conductor and stable under reducing conditions.[37], [38] These properties make $SrVO_3$ a promising electron emission cathode material worthy of further study. This naturally leads to the question: what other perovskite oxides might also have a low work function, good electrical conductivity, and good chemical stability, and thus be promising electron emission materials?

Motivated by these questions of materials discovery and design of low work function perovskites, we performed high-throughput screening using DFT methods to assess the predicted work function, thermodynamic stability under operating conditions, and conduction characteristics of 2913 perovskite oxide compositions. Our high-throughput screening resulted in the discovery of seven compounds predicted to have low (< 2.5 eV) AO-terminated (001) surface work functions. Examples of promising compounds include $BaMoO_3$, $SrNb_{0.75}Co_{0.25}O_3$, $BaZr_{0.375}Ta_{0.5}Fe_{0.125}O_3$ and $BaZr_{0.375}Nb_{0.5}Fe_{0.125}O_3$, which are predicted to have work functions of 1.1 eV, 1.5 eV, 0.9 eV and 1.6 eV, respectively. In addition, the new HSE-level work function data obtained from this screening allowed us to further verify and refine the correlation between surface work function and bulk O *p*-band center first proposed in [33]. Finally, the large amount of data obtained from this computational search enabled an enhanced understanding of the materials chemistry governing the work function of perovskite oxides. Specifically, we discovered a qualitative empirical relationship between the work function and the number of *d* electrons, where materials with barely filled *d* bands (e.g. only one or two *d* electrons) result in the lowest work functions while insulators containing zero *d* electrons and late transition metal perovskites with nearly filled *d* bands result in the highest work functions.

## 2 Results and Discussion

### 2.1 High-throughput materials screening procedure and elimination criteria

**Figure 1** contains an overview of the high-throughput screening process and elimination criteria used in this work. The complete set of 2913 perovskite materials was obtained in part from a previous high-throughput screening study of perovskite oxides for the application of stable, high activity oxygen reduction catalysts for solid oxide fuel cells (the full list of materials can be found in **Supplementary Information Section 6**).[39] The structure of perovskite oxides is shown in **Figure 2(a)**. The remaining perovskite materials were simulated in this study with the aid of the MAterials Simulation Toolkit (MAST).[40] Additional details of the composition spaces explored and other computational details can be found in **Section 4**. Here, we list our chosen elimination criteria used to successively reduce the space of promising perovskite materials for use as low work function electron emitters. The subsequent sections provide our rationale for the choice of each of these elimination criteria, together with the results of the screening analysis for each criterion. The first screening step used the bulk O $p$-band center as a descriptor of the work function to eliminate materials with a predicted work function higher than 2.01 eV, which is the predicted value of $SrVO_3$ at the GGA level. The second screening step used this down-selected pool of low predicted work function candidates further eliminated compounds expected to be less stable under typical thermionic cathode operating conditions than $SrVO_3$ based on convex hull analysis, i.e. materials with an energy above the convex hull greater than 43 meV/atom, the calculated stability value of $SrVO_3$ at the GGA+$U$ level. We note that this screening prioritizes work function, as that is our main interest. However, this approach uses a rather approximate value (the work function prediction) in the first screening step, and it is possible that screening on stability would be a better first step, followed by work function. In practice we perform all the necessary bulk DFT calculations for both properties together and screen on both together, so the order has no impact. The third screening step assessed the DFT-predicted electronic conductivities of the pool of predicted low work function, stable materials using HSE functional. In our screening process we eliminated materials that were predicted to have bandgaps of 0.5 eV or greater, based on the expectation that they will be poor electrical conductors. Finally, the last screening step is to directly calculate the (001) AO-terminated work function at the HSE level and eliminate compounds with a work function of greater than 2.5 eV, therefore producing the set of new promising perovskite materials as low work function thermionic cathodes. We remark here that materials that are insulating but have low work function, while not promising for use as thermionic cathodes in high power VED applications, may still find use as field emission coatings. For example, the materials $Ba_{0.65}Sr_{0.35}TiO_3$[41], [42] and N-doped $SrTiO_3$[43], [44] have been successfully used as thin film, work-function-lowering field emission coatings on Si, as well as direct field emission from nanotips of $PbZr_xTi_{1-x}O_3$.[45] As the field emission process relies on a low work function and electron tunneling through the cathode surface, the use of an insulating material is less of a critical design issue.

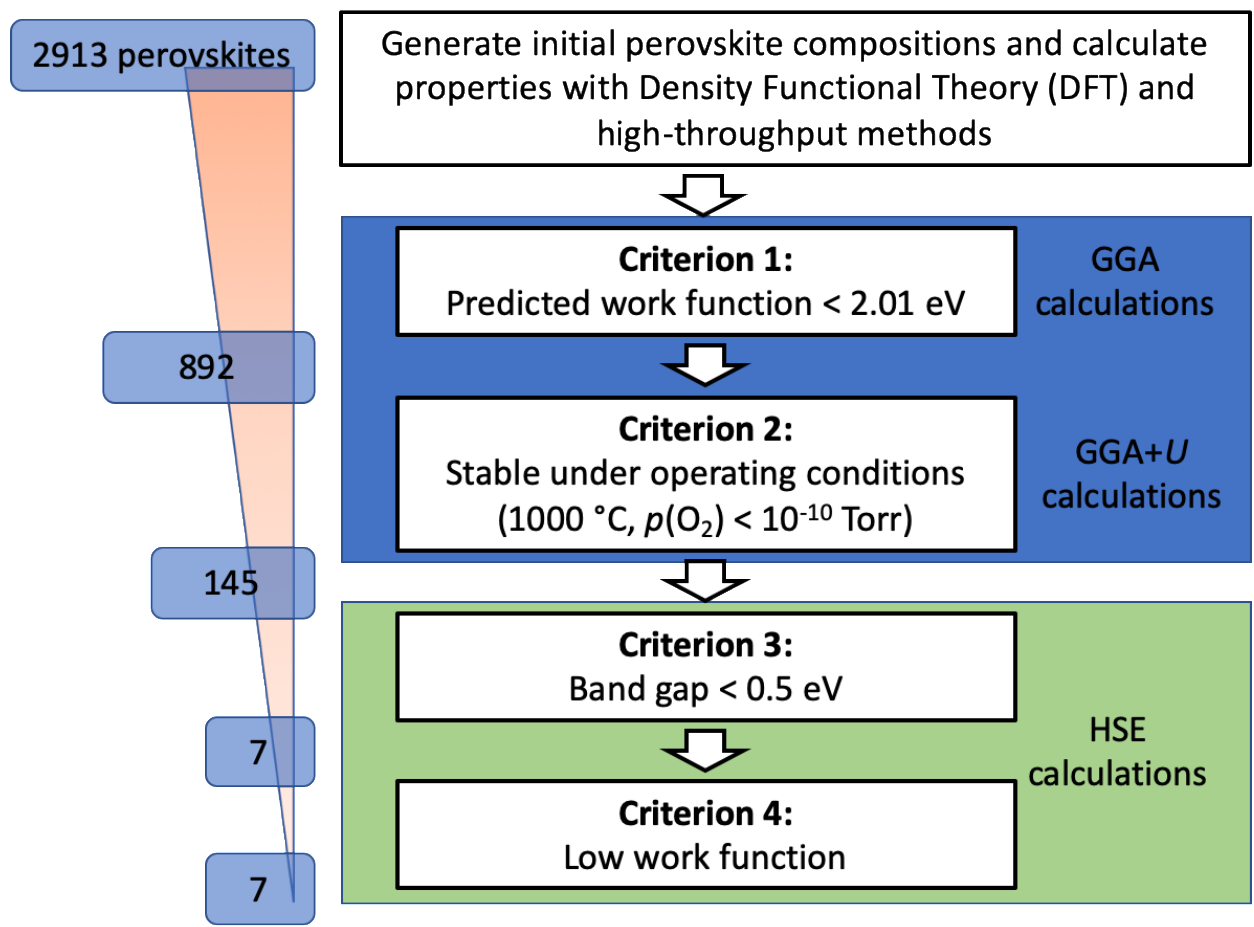

**Figure 1.** Diagram illustrating the DFT-based high-throughput screening process used in this work. At each step of the screening process, an elimination criterion is invoked to reduce the pool of potentially promising perovskite compounds.

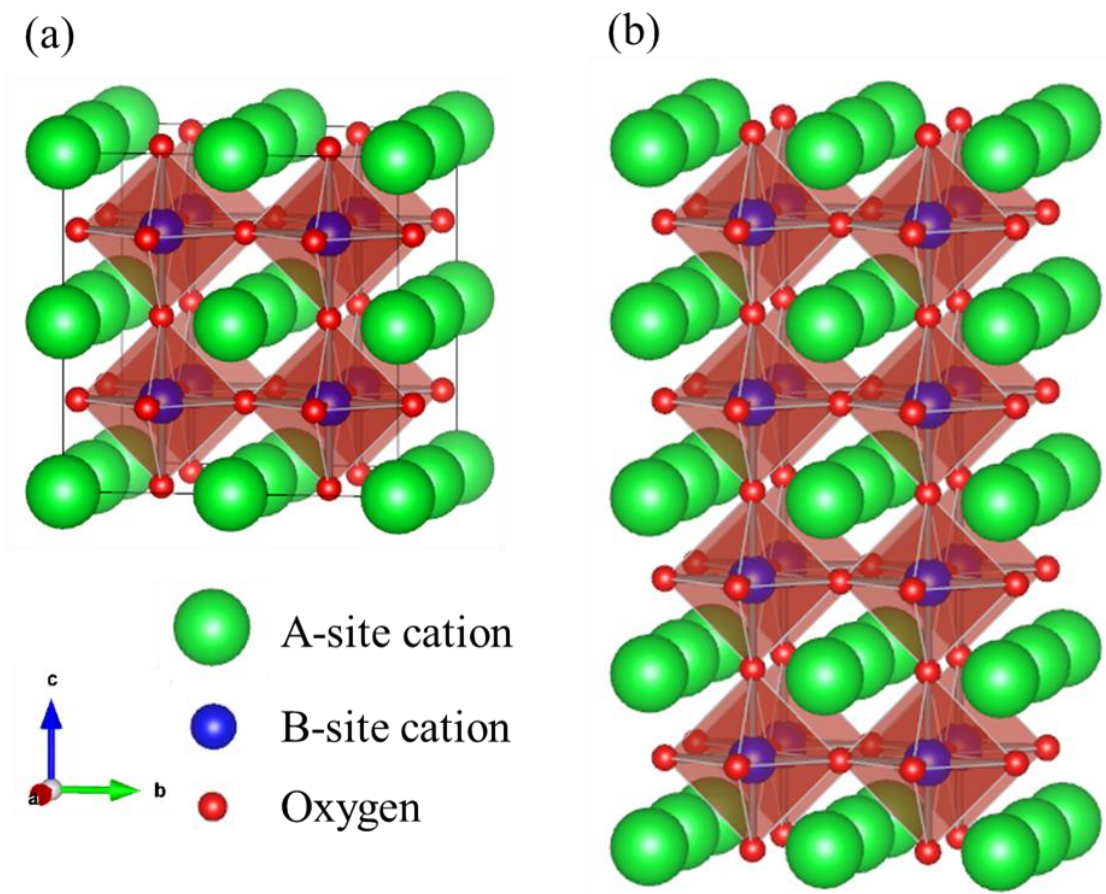

**Figure 2.** (a) Structure of typical perovskite oxides. (b) Structure of (001)-AO terminated perovskite surface.

### 2.1.1 Elimination criterion 1 (predicted work function)

Reference [33] demonstrated a linear relationship between the HSE-calculated work function (for both AO- and $BO_2$-terminated (001) surfaces) and bulk HSE O $p$-band center. However, it was computationally intractable to screen the full set of 2913 perovskites in this study at the HSE level. Therefore, it was first necessary to verify this correlation persists at the GGA level. Here, we chose to evaluate the O $p$-band center trend at the GGA level as opposed to the GGA+$U$ level based on recent findings that showed that GGA resulted in improved statistical correlations of numerous perovskite properties with the O $p$-band center compared to the use of GGA+$U$.[46] **Figure 3** contains a plot of the GGA-calculated AO work function of a representative set of 27 perovskites as a function of the bulk GGA O $p$-band center. We note here that in this work, all predicted and calculated work functions are for (001) oriented perovskites with the AO surface termination, whose structure is shown in **Figure 2(b)**. This was chosen as the surface of interest based on work reported in [33], [34] that showed that these surfaces tend to have low work functions due to surface dipoles directed away from the terminating surface that lower the electrostatic barrier and the work function. The set of perovskites used to construct the correlation in **Figure 3** includes: La(Sc, V, Cr, Mn, Fe, Ru, Co, Ni, Cu, Al, Ga)$O_3$, Sr(Ti, Zr, V, Nb, Cr, Mo, Mn, Fe, Ru, Co)$O_3$, and Ba(Ti, Zr, Hf, Mo, Nb, Sn)$O_3$. **Figure 3** demonstrates there is a linear relationship between the GGA AO work function and GGA bulk O $p$-band center with an $R^2$ value of 0.83, this correlation is slightly higher although generally quite similar to that of AO work functions for metallic perovskites at the HSE level in [33] ($R^2 = 0.77$). Although the largest work function spread in this correlation is more than 1 eV, the RMSE of this fit is only 0.29 eV. By applying the correlation between work function and O $p$-band center in **Figure 3**, we eliminated candidate materials with predicted work function higher than 2.01 eV, the predicted work function of $SrVO_3$ at the GGA level. $SrVO_3$ was chosen as the reference system to ascertain whether other perovskites have a sufficiently low predicted work function to be interesting for two reasons: (1) $SrVO_3$ has already been simulated at the HSE level and shown to have a low (001) AO terminated HSE work function of about 1.9 eV,[33] thus making it a promising material, and (2) our choice of using a known promising material for the work function cutoff is a conservative one, which will likely minimize the number of false positive predictions. Based on the reference work function value of 2.01 eV and the RMSE of 0.29 eV, it is most likely that the resulting materials will have work functions lower than 2.59 eV, which is two standard deviations above the cutoff. Given the fact that $LaB_6$, a widely used commercial cathode, has a work function of 2.6 eV,[47] our identified low work function candidates will likely have comparable or lower work functions than $LaB_6$ and thereby justify further, in-depth experimental assessment.

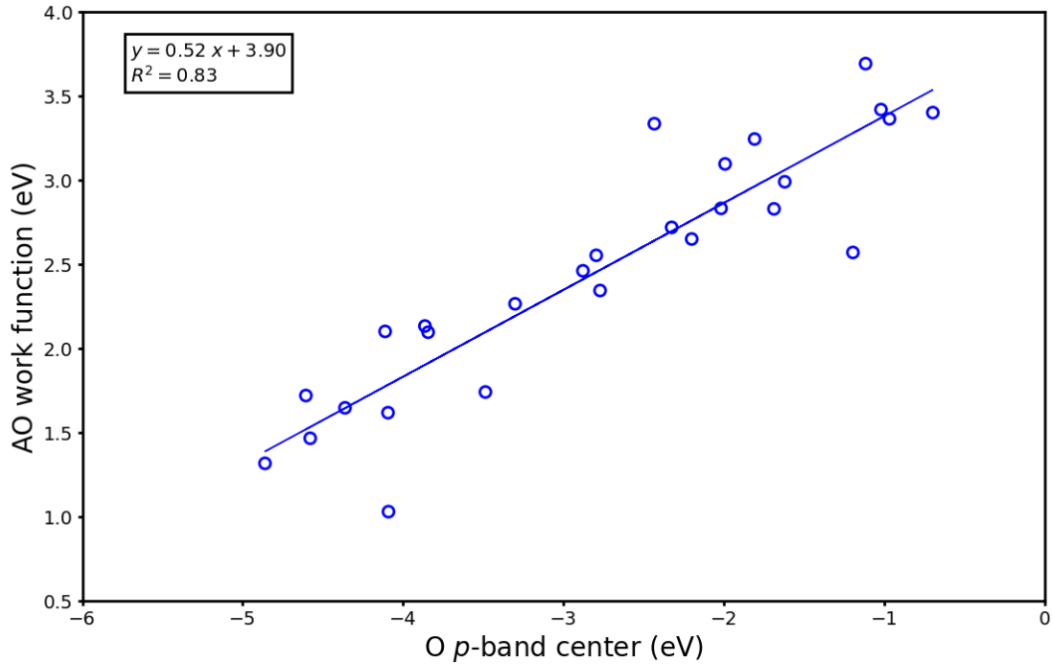

**Figure 3.** Calculated GGA AO surface work functions of a representative set of 27 different perovskites with respect to their bulk GGA O *p*-band centers.

The results in **Figure 3** were fit to line to yield the following relationship between the GGA (001) AO-terminated work function and the bulk O *p*-band center:

$$\Phi_{AO,pred}^{GGA} = 0.52 \times O_{p-band} + 3.90 \qquad (1)$$

Using this relationship between predicted GGA work function and O *p*-band center, we calculated the predicted work function of all 2913 simulated perovskites. In this screening step, 892 perovskites have a predicted work function lower than the SrVO$_3$ predicted GGA value of 2.01 eV.

**2.1.2 Elimination criterion 2 (stability under operating conditions)**

Next, we chose to exclude perovskites that are calculated to be unstable in a typical thermionic cathode operating environment of high temperature and ultra-high vacuum corresponding to $T = 1000$ °C and $p(O_2) = 10^{-10}$ Torr. We calculated the energy above the convex hull ($E_{hull}$) with GGA+$U$ as a measure of material stability. $E_{hull}$ measures the decomposition energy of a material into the set of most stable phases, with a value of zero indicating the material under investigation resides on the convex hull and is formally stable. Recent studies have shown that there are numerous materials that are calculated from DFT to reside above the convex hull, yet can be routinely synthesized and used in functional applications. These studies have found that a more realistic cutoff of material stability may be up to about 50 meV/atom above the hull.[48]–[50] From our DFT calculations, we found that SrVO$_3$ has an $E_{hull}$ value of 42 meV/atom under thermionic cathode operating conditions (see **Section 4** for details of how the energy above convex hull was calculated). This value is within the $E_{hull}$ range of materials found to be readily synthesizable, and SrVO$_3$ has been experimentally made and used in applications ranging from oxide electronics to anodes for solid oxide fuel cells.[37], [51]–[53] Therefore, to focus on a smaller set of materials with even higher probability of good stability, we used the $E_{hull}$ value of 42 meV/atom as the stability elimination criterion.

**Figure 4** contains a plot of the predicted work function (calculated from the O *p*-band center descriptor trend in **Figure 3**) as a function of the calculated $E_{hull}$ values for all materials analyzed in this study. In the scatterplot of **Figure 4**, the data are grouped according to the number of *d*-band electrons, as determined by electron counting rules assuming completely ionic bonding (e.g. for SrVO$_3$, V is in the 4+ oxidation state, so this is a 3$d^1$ material, and the *d*-band occupation is assumed to be 1) (see **Section 4** for more information on how the *d*-band electron counts were calculated). **Figure 4** illustrates that materials with zero *d*-band electrons have high predicted work functions while materials with a low, but nonzero, number of *d*-band electrons (greater than zero but less than or equal to three) have the lowest predicted work functions. Materials with *d*-band electron counts greater than three but less than or equal to ten have a spread in work function values, but their predicted work functions are similar to materials with zero *d*-band electrons, i.e., consistently higher than the materials with few (but nonzero) *d*-band electrons. These trends are discussed in more detail in **Section 2.3**. Applying our stability screening, the previous pool of 892 candidate materials was reduced to 145 materials.

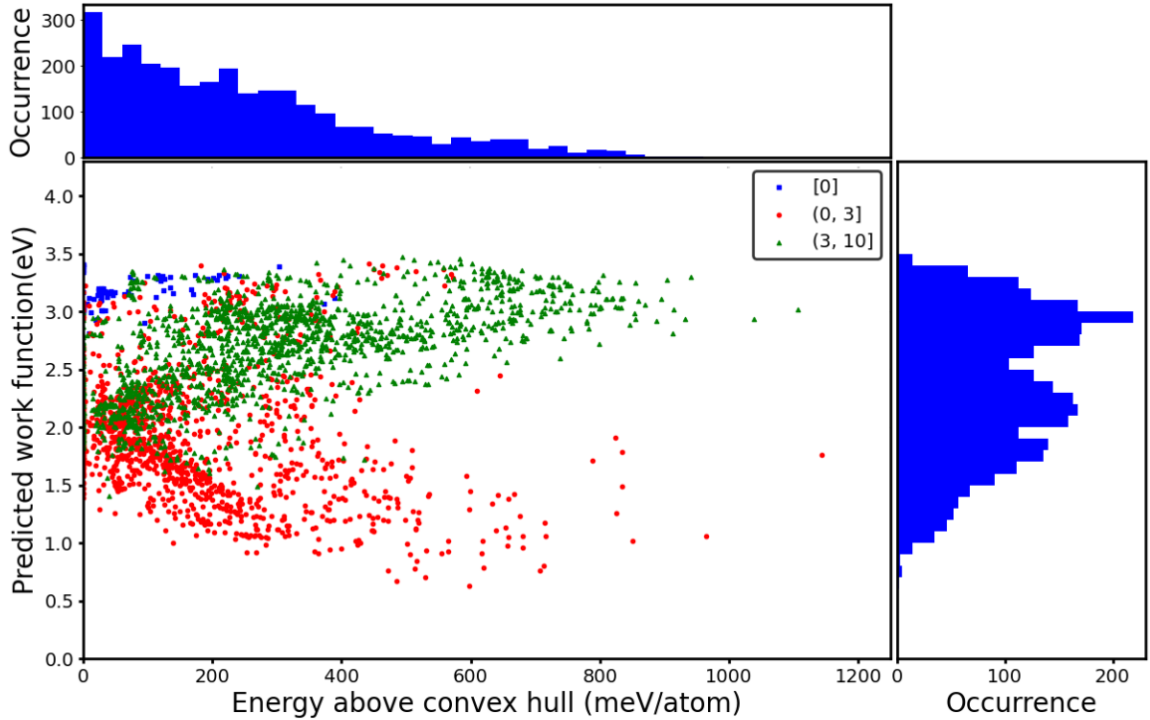

**Figure 4.** Plot of predicted work functions calculated using the O $p$-band center descriptor as a function of $E_{hull}$ under typical thermionic cathode operating conditions for all materials in this study. The two histograms along $x$- and $y$-axis demonstrate the distribution of energies above convex hull and the distribution of predicted work functions, respectively. The blue, red and green symbols denote perovskites with $n$ electrons in the $d$-band ($n = 0$, $0 < n \leqslant 3$, $3 < n \leqslant 10$, respectively), based on electron counting rules assuming completely ionic bonding.

### 2.1.3 Elimination criterion 3 (electrical conductivity)

The third step of our screening process was to remove candidate materials that are unlikely to be sufficiently electrically conductive for use as thermionic electron emitters. It is necessary for a thermionic emitter material to be electrically conductive in order to realize the desired high current emission densities required in modern vacuum electronic devices. We focus on materials with low band gap, because for small band gap materials there will be many charge carriers excited into the conduction band leading to large conductivity. The opposite will occur for a large band gap material. We note that this approach does not screen for more subtle drivers of low conductivity, e.g., large polaron hopping barriers, which are beyond the scope of this work. We assumed that only examining materials with zero bandgap was too stringent a cutoff. The perovskites examined in this work typically contain transition metals in high oxidation states (e.g. $Co^{4+}$). Consequently, additional physics of these materials may occur at the high operating temperatures of thermionic cathodes that could produce a reduction in the bandgap compared to the zero temperature DFT-calculated HSE bandgap. These physics could include: magnetic transitions[54], formation of oxygen vacancies[55], or small structural changes (e.g. Jahn-Teller distortion[54]) that would influence the electronic structure without resulting in a large structural change but were not explicitly included in our simulations. While it is difficult to quantify the energy scale of these effects on the band gap in a general way, we assumed a reasonable upper bound of 0.5 eV. Therefore, we used a band gap of 0.5 eV as the threshold value, and eliminated compounds with an HSE-calculated band gap greater than 0.5 eV. As aforementioned, such insulating materials may find use in field emission applications, but are not in the scope of this study.

145 materials passed the predicted work function and stability elimination criteria and a summary of the key calculated properties of these materials is provided in the **Supplementary Information**. In this screening step, we calculated the bulk HSE-level band gaps of 25 of these 145 compounds which have the lowest predicted work functions according to Eq. (1), based on their O $p$-band centers. The results are tabulated in **Table 1**. Based on this analysis of 25 compounds, 4 compounds were predicted to have band gaps less than 0.5 eV while the remaining 21 compounds were predicted to have band gaps higher than 0.5 eV, or insulating, and thus not useful as vacuum electronic cathodes. We therefore chose to focus our remaining analyses on the four low band gap, (probably) conductive compounds: $Sr_{0.5}La_{0.5}MoO_3$, $Ba_{0.25}Ca_{0.75}MoO_3$, $SrMoO_3$ and $BaZr_{0.375}Ta_{0.5}Fe_{0.125}O_3$.

**Table 1.** Summary of the predicted work functions, calculated stabilities, and HSE band gaps of perovskites passing the first two elimination criteria.

| Composition | Predicted work function (eV) | $E_{hull}$ (meV/atom) | HSE band gap (eV) |
|---|---|---|---|
| $La_{0.75}Y_{0.25}MoO_3$ | 1.29 | 25 | 1.47 |
| $Ca_{0.25}La_{0.75}MoO_3$ | 1.39 | 0 | 0.66 |
| $LaMn_{0.125}Ga_{0.875}O_3$ | 1.41 | 41 | 3.25 |
| $La_{0.5}Y_{0.5}MoO_3$ | 1.42 | 0 | 1.51 |
| $La_{0.75}Gd_{0.25}VO_3$ | 1.43 | 0 | 2.20 |
| $Sr_{0.5}La_{0.5}MoO_3$ | 1.43 | 0 | 0.47 |
| $Gd_{0.5}Y_{0.5}MoO_3$ | 1.45 | 24 | 1.48 |
| $Ca_{0.5}La_{0.5}MoO_3$ | 1.47 | 0 | 1.16 |
| $La_{0.5}Gd_{0.5}VO_3$ | 1.47 | 0 | 2.18 |
| $LaTi_{0.5}Fe_{0.5}O_3$ | 1.50 | 31 | 2.95 |
| $LaMg_{0.125}V_{0.875}O_3$ | 1.51 | 0 | 1.33 |
| $Pr_{0.5}Gd_{0.5}VO_3$ | 1.52 | 0 | 2.18 |
| $Pr_{0.5}Gd_{0.5}MoO_3$ | 1.52 | 0 | 1.40 |
| $LaV_{0.5}Fe_{0.5}O_3$ | 1.56 | 38 | 2.01 |
| $YMoO_3$ | 1.58 | 0 | 1.71 |
| $Pr_{0.75}Gd_{0.25}VO_3$ | 1.58 | 0 | 2.19 |
| $CaMoO_3$ | 1.58 | 0 | 0.55 |
| $Ba_{0.25}Ca_{0.75}MoO_3$ | 1.59 | 14 | 0.02 |
| $Gd_{0.25}Y_{0.75}MoO_3$ | 1.59 | 0 | 1.36 |
| $SrMoO_3$ | 1.59 | 0 | 0.03 |
| $GdVO_3$ | 1.60 | 25 | 2.26 |
| $Gd_{0.5}Y_{0.5}VO_3$ | 1.60 | 0 | 2.28 |
| $LaMg_{0.25}V_{0.75}O_3$ | 1.61 | 0 | 1.39 |
| $BaZr_{0.375}Ta_{0.5}Fe_{0.125}O_3$ | 1.62 | 41 | 0.38 |
| $LaCrO_3$ | 1.63 | 0 | 0.76 |

**2.1.4 Elimination criterion 4 (quantitative work function calculation)**

**Table 2** summarizes the $E_{hull}$, HSE-calculated bandgaps, and HSE-calculated work functions, for the four materials obtained from the bandgap screening step. In addition, it includes another three materials, outside of the top 25 predicted low work function materials listed in **Table 1**. These three materials were from an earlier version of our screening analysis, where we used a linear correlation between O $p$-band center and work function at the GGA+$U$ level (criterion 1). This correlation was not as accurate as that at the GGA level, so we did not use it in the final analysis in this paper. However, based on this earlier approach, these three perovskites passed our first three elimination criteria (see Figure 1) and were found to be promising. We included them in **Table 2** for completeness. Our calculations show that for these seven materials, all of them have HSE-calculated work functions below 2.5 eV. These materials include $BaMoO_3$ and $BaZr_{0.375}Ta_{0.5}Fe_{0.125}O_3$, which have very low work functions of 1.1 eV and 0.9 eV, respectively. The cells used in these calculations have not included any effort to capture surface segregation.

Among these perovskites in **Table 2**, $Ba_{0.25}Ca_{0.75}MoO_3$ might be expected to have strong Ba segregation due to its larger radius than Ca. Although such segregation is not included in our calculations, it would likely increase the surface dipole and reduce the work function since Ba has a lower electronegativity and larger size than Ca. Therefore, segregation would likely increase the possibility that this is a material of interest. For $Sr_{0.5}La_{0.5}MoO_3$, Sr and La have similar atomic radius and might expect Sr segregation as has been found in some other perovskites. However, the balancing of size, charge, and dipole effect on such surfaces is highly complex, making it difficult to determine the correct surface segregation[56] and detailed studies to explore this surface are beyond the scope of this work. The other five perovskites are expected to have no significant segregation on (001)-AO termination since they only have one A-site cation. Overall, we have found seven perovskite materials that emerge as promising, new, stable, low work function, conductive materials for further exploration as thermionic electron emitters.

**Table 2.** Summary of HSE-calculated work functions for all materials passing elimination criteria 1-3, and the most promising compounds identified in this work.

| Composition | $E_{hull}$ (meV/atom) | HSE bandgap (eV) | HSE work function (eV) |
|---|---|---|---|
| $SrMoO_3$ | 0 | 0.03 | 1.93 |
| $BaMoO_3$ | 24 | 0.38 | 1.06 |
| $Sr_{0.5}La_{0.5}MoO_3$ | 0 | 0.47 | 1.86 |
| $Ba_{0.25}Ca_{0.75}MoO_3$ | 14 | 0.02 | 1.68 |
| $SrNb_{0.75}Co_{0.25}O_3$ | 37 | 0.03 | 1.51 |
| $BaZr_{0.375}Ta_{0.5}Fe_{0.125}O_3$ | 41 | 0.38 | 0.93 |
| $BaZr_{0.375}Nb_{0.5}Fe_{0.125}O_3$ | 21 | 0.41 | 1.56 |

**2.2 Refinement and validation of O *p*-band center as a work function descriptor**

We have used these seven work function values in **Table 2** and the original 20 work function values from [33] to refine the trends of HSE-calculated AO surface work functions versus the bulk HSE O *p*-band center previously revealed in [33]. In addition, as a further test of this correlation, we chose a select group of perovskites found to have very low O *p*-band centers that are lower in value than the materials comprising the original trend, and also included the calculated HSE work functions of these materials. These test materials were chosen solely based on their low O *p*-band center values, regardless of their stability or band gap values. These additional test materials consisted of: $BaNbO_3$, $SrNbO_3$, $PrHfO_3$, $BaTaO_3$, $Ca_{0.5}Ba_{0.5}Ta_{0.5}Nb_{0.5}O_3$, $La_{0.5}Ba_{0.5}Ta_{0.5}Nb_{0.5}O_3$, $La_{0.5}Ba_{0.5}NbO_3$, $Ba_{0.5}La_{0.25}Rb_{0.25}NbO_3$ ($La_{0.5}Ba_{0.5}O$- and $La_{0.5}Ba_{0.25}Rb_{0.25}O$-terminated AO surfaces), and $La_{0.5}Rb_{0.5}NbO_3$.

**Figure 5** shows the HSE-calculated AO surface work function versus the HSE bulk O *p*-band center values for materials from [33], the promising materials from the screening conducted in this work, as well as the above-discussed test set of very low predicted work function materials. In comparison to the trend of GGA work function and O *p*-band center values in **Figure 3**, the linear correlation between work function and O *p*-band center in **Figure 5** splits into two trend lines. Consistent with the results and explanation from [33], these two distinct trendlines are the result of differences in the electronic structure of these two sets of materials. The materials comprising the top trend line are either band or Mott-Hubbard insulators with a nonzero HSE-calculated bandgap and the materials comprising the bottom trend line are correlated metals.[33]

Now that many more perovskite materials have been included in the trends displayed in **Figure 5**, some comparisons can be drawn between the trends in **Figure 5** and the original trends presented in [33]. In [33], the insulator and metal trend lines had $R^2$ values of 0.78 and 0.77, respectively. In **Figure 5**, the updated lines of best fit show the insulator and metal trend lines have $R^2$ values of 0.47 and 0.80, respectively. It is evident the inclusion of more data results in a significantly worse correlation for insulating compounds, but a similar correlation for metallic compounds. Regarding insulating compounds, the poor correlation is expected as the calculated work functions for (001) AO surfaces of insulating perovskites are limited to a small range between about 2.5-3.5 eV so even a spread in the work function values of a few tenths of an eV relative to the trend line results in a low $R^2$ value. Regarding metallic compounds, the consistent correlation despite our inclusion of new data spanning a larger domain of O *p*-band center values and a larger dynamic range of (001) AO work functions between about 0.8-3.3 eV is encouraging. For the metallic compounds, the corresponding RMSE/$\sigma$ = 0.45, much smaller than the insulator case (RMSE/$\sigma$ = 0.73), indicating a statistically more robust fit to the metallic compound work functions.

Although the O *p*-band center is a reasonably accurate descriptor of HSE work functions for metallic perovskites, one disadvantage of employing this bulk descriptor is it cannot capture the effect of different surface dipoles yielding different work functions for a particular material. As a concrete example, **Figure 5** shows the work functions of $Ba_{0.5}La_{0.25}Rb_{0.25}NbO_3$ with two different surface terminations: the $Ba_{0.5}La_{0.5}O$ surface (work function of 0.99 eV) and the $Ba_{0.5}La_{0.25}Rb_{0.25}O$ surface (work function of 0.79 eV). $Ba_{0.5}La_{0.25}Rb_{0.25}NbO_3$ has a single bulk O *p*-band center value, yet has AO work functions that differ by 0.2 eV. Calculations by Lee *et al.* examining the trends of oxygen reduction and intermediate species' binding energies on perovskite (001) AO and $BO_2$ surfaces using the O *p*-band center demonstrated that the use of the subsurface O *p*-band center (instead of the bulk value) could resolve the binding energy difference of the AO and $BO_2$ surfaces for a single perovskite material.[57] A similar approach used to discern work function differences arising from differences in surface dipoles may work for the materials investigated here, but doing so would lose the advantage of using a bulk descriptor to enable accelerated predictions of work function.

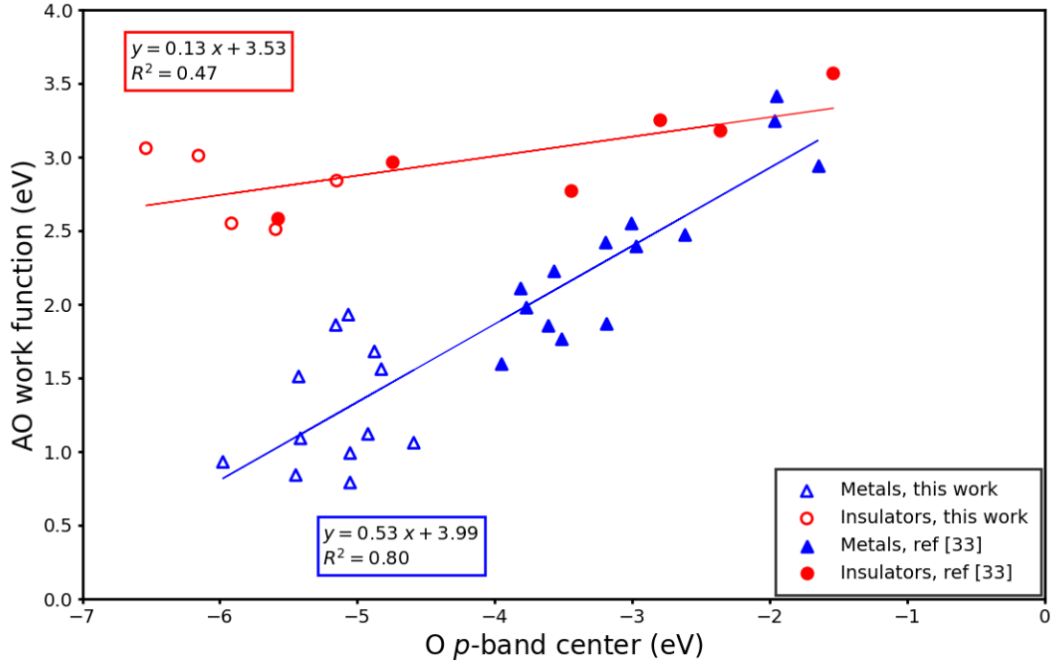

**Figure 5.** Refined correlation of the HSE (001) AO-terminated work functions as a function of bulk HSE O *p*-band center including original data from the work of Jacobs *et al.* (filled red and blue circles),[33] promising materials identified from the screening procedure employed in this work (empty red and blue triangles), and select tests of very low O *p*-band center materials predicted from elimination criterion 1 in this work (empty red and blue squares).

## 2.3 Materials electronic structure principles to rationalize low and high work function perovskites

Based on the analysis of predicted work function and stability discussed in **Section 2.1.2** and shown qualitatively in **Figure 4**, we determined that the number of *d*-band electrons in these perovskites has a significant impact on determining their work functions. Materials with zero *d*-band electrons have a high predicted work function, materials with a low number of *d*-band electrons (less than three but greater than zero) have the lowest work functions, while materials with higher number of *d*-band electrons (greater than three and less than a full *d*-band of 10) also have high work functions, in the same range as the materials with zero *d*-band electrons. In **Figure 6** we have taken all of the data from **Figure 4** and produced a bar chart histogram of the fraction of perovskites with a particular predicted work function, grouped by the number of *d*-band electrons. **Figure 6**, shows that materials with zero *d*-band electrons, have the highest average predicted work function of 3.17 eV with a standard deviation 0.13 eV. Materials with more than zero but less than or equal to three *d*-band electrons have an average predicted work function of 1.92 eV with a standard deviation of 0.54 eV, Materials with more than three but less than or equal to ten *d*-band electrons have an average predicted work function of 2.67 eV with a standard deviation of 0.41 eV, which is qualitatively similar to the results for materials with zero *d*-band electrons. These trends in **Figure 6** agree with the electronic structure and bonding trends discussed in the work of [33], which discussed the physical basis for these trends from the interplay of bonding ionicity, O *p*-band center position, and the resulting work function value for different types of perovskites. For example, $LaScO_3$ with 0 *d*-band electrons has a high work function, $SrVO_3$ with few *d*-band electrons has a low work function, and $LaNiO_3$ with a nearly full *d*-band also has a high work function.[33]

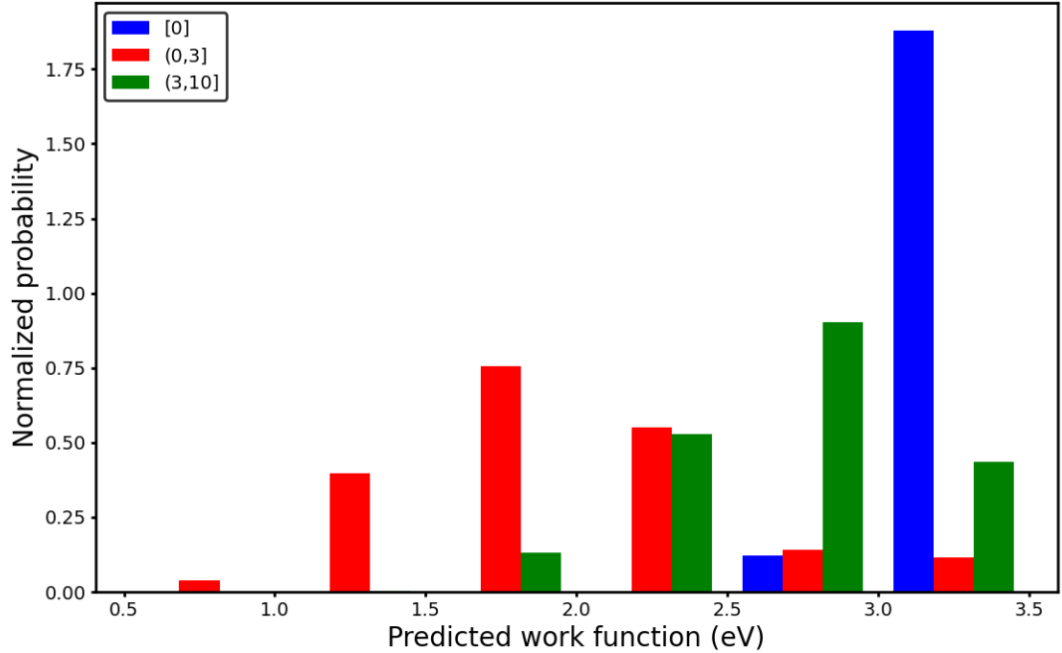

**Figure 6.** Bar chart histogram showing the distribution of predicted work functions, grouped by the number of *d*-band electrons.

## 2.4 Engineering SrVO₃ stability and work function through alloying

From the dataset established during our screening, we extract the stabilities and predicted work functions of alloyed SrVO$_3$ and summarize how different alloy elements affect the properties of SrVO$_3$ in this section. We chose to study the effect of alloying on SrVO$_3$ since SrVO$_3$ has been extensively studied previously [38, 40] and is under current experimental investigation.[58] It therefore provides a convenient reference material for additional refined investigations, such as these alloying effects.

**Figure 7** contains bar plots showing the effect of alloying on the stability and predicted work function when various elements are alloyed on the A-site, B-site, and combined A- and B-sites of SrVO$_3$. From our DFT calculations, the addition of Nb and Ta on the B-site for a range of compositions is expected to enhance the stability, and the alloying of La on the A-site, Nb and Ta on the B-site, and combined alloying of La/Nb and La/Ta on the A-/B-sites is expected to reduce the work function. Overall, these results suggest that the alloyed systems SrV$_{1-x}$Nb$_x$O$_3$, SrV$_{1-x}$Ta$_x$O$_3$, as well as these systems doped with La, deserve further exploration as stable, low work function perovskites. We note that recent studies have shown that SrV$_{1-x}$Nb$_x$O$_3$ alloys with different Nb content can be synthesized, with promising application as anodes for solid oxide fuel cells.[53], [59]

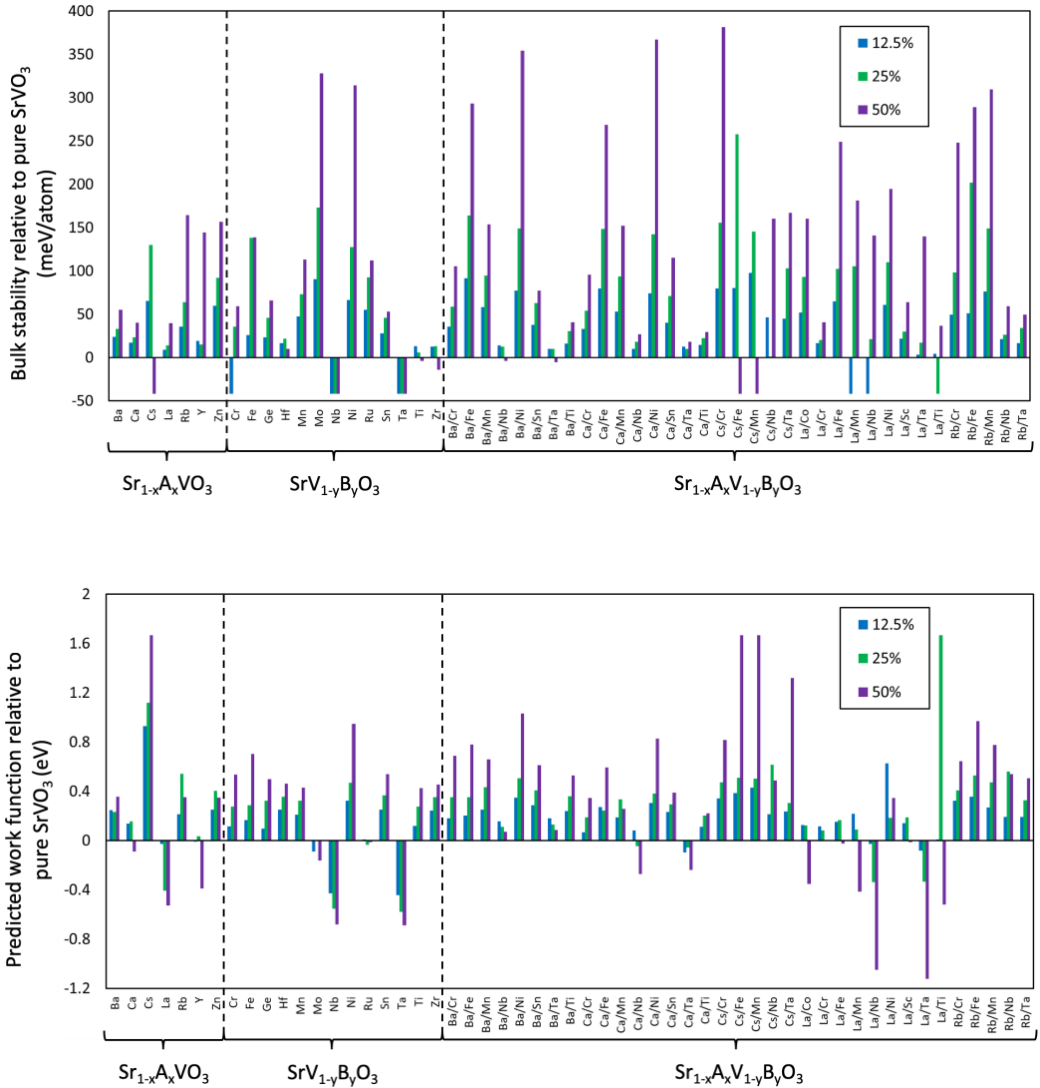

**Figure 7.** Effect of stability and predicted work function from alloying SrVO$_3$ with various dopants on the A-site (e.g. Sr$_{1-x}$A$_x$VO$_3$), the B-site (e.g. SrV$_{1-y}$B$_y$O$_3$) and both the A- and B-site together (e.g. Sr$_{1-x}$A$_x$V$_{1-y}$B$_y$O$_3$). In both plots, the blue, green and purple bars denote site fraction alloying of 12.5%, 25% and 50%, respectively. (A) and (B): alloying effect on stability and work function of SrVO$_3$, respectively. The values of stability and predicted work function are given relative to the pure compound, therefore negative values indicate enhanced stability and reduced work function relative to the unalloyed material.

## 3 Summary and Conclusions

In this study, our goal was to discover new stable, low work function perovskite compounds for use as thermionic electron emission cathodes using high-throughput DFT methods. From a pool of 2913 compounds, we successively down-selected sets of materials that were found have a low predicted work function, be stable under typical thermionic cathode operating conditions, be sufficiently conductive by virtue of a small electronic bandgap, and finally, possess a low work function (001) AO-terminated surface. From our initial pool of about 2913 compounds, 145 compounds were found to have a low predicted work function and be stable under operating conditions. Of these 145 materials, seven of them were found to be sufficiently conductive and have a low work function. These promising materials consist of $SrMoO_3$, $BaMoO_3$, $Sr_{0.5}La_{0.5}MoO_3$, $Ba_{0.25}Ca_{0.75}MoO_3$, $SrNb_{0.75}Co_{0.25}O_3$, $BaZr_{0.375}Ta_{0.5}Fe_{0.125}O_3$, and $BaZr_{0.375}Nb_{0.5}Fe_{0.125}O_3$. These new compounds, together with $SrVO_3$, constitute the set of most promising materials for new thermionic electron emission cathodes for use in high frequency, high power vacuum electronic devices and potentially as emissive layers in thermionic and photon-enhanced thermionic energy converters.

The large amount of perovskite data analyzed in this work enabled the understanding of what qualitatively governs the work function of a perovskite based on simple empirical electron counting rules. Here, we found that materials with a small number of electrons (e.g. 1 to 3) in the $d$-band tend to result in the lowest predicted work functions, while insulating compounds with 0 $d$-band electrons and materials containing late transition metals with highly occupied $d$-bands tend to result in much higher predicted work functions. These trends in materials chemistry provide a simple foundation for qualitatively identifying whether particular perovskite compounds may be expected to have a low or high work function value for a given application.

This work provides a basis for further computational and experimental study of perovskite work functions. Computationally, further study of the promising materials discovered here may involve assessing effects on the stability and work function, such as: cation and oxygen off-stoichiometry, the stability and work function of different surface orientations with and without adsorbed species, and high temperature perovskite structural transitions (e.g. orthorhombic to cubic transitions). Experimentally, the synthesis, characterization, work function measurement and high temperature emission properties of the promising materials discovered here is critical to assessing the quality of these materials as thermionic electron emitters. Finally, the expanded data set of HSE-level perovskite work function calculations in this study may find use in design of perovskite heterojunctions where knowledge of electronic band offsets and dipole engineering approaches are critical.

## 4. Computational methods

We performed all simulations using Density Functional Theory (DFT) as implemented in the Vienna Ab Initio Simulation Package (VASP).[60] Perdew-Burke-Ernzerhof (PBE)[61]-type pseudopotentials utilizing the projector augmented wave (PAW)[62] method were used for all atoms. All calculations were performed with spin polarization enabled. The planewave cutoff energy was set to 500 eV. A $4 \times 4 \times 4$ Monkhorst-Pack[63] $k$-point mesh was used for bulk GGA and GGA+$U$[64] relaxations and a $4 \times 4 \times 1$ Monkhorst-Pack $k$-point mesh was used for surface slab simulations at the GGA level. A $2 \times 2 \times 2$ Monkhorst-Pack $k$-point mesh was used for bulk HSE[36] relaxations and a $2 \times 2 \times 1$ Monkhorst-Pack $k$-point mesh was used for surface slab simulations at the HSE level. For the GGA+$U$ simulations, the $U$ values of V, Cr, Mo, W, Mn, Fe, Co, Ni were 3.25 eV, 3.7 eV, 4.38 eV, 6.2 eV, 3.9 eV, 5.3 eV, 3.32 eV, and 6.2 eV,[65] respectively, and were chosen to maintain consistency with the Materials Project database, which is necessary for accurate thermodynamic stability calculations.[65] For HSE calculations, the fractions of exact exchange were obtained from previous studies.[33], [66], [67] In these previous studies, the fraction of exact exchange was chosen to reproduce the experimental bandgaps and densities of states, and were chosen in these previous studies as 0.25 (LaScO$_3$), 0.15 (LaTiO$_3$, LaCrO$_3$, LaMnO$_3$, LaFeO$_3$), 0.125 (LaVO$_3$), 0.05 (LaCoO$_3$), and 0 (LaNiO$_3$).[66], [67] For the materials SrVO$_3$, CaNbO$_3$, SrNbO$_3$, and pure and doped BaNbO$_3$, an exact exchange value of 0.125 was used. This value of 0.125 was used because these V- and Nb-based compounds are in the same column of the periodic table and thus are expected to be chemically similar to LaVO$_3$, for which an exact exchange value of 0.125 was found to result in a good fit with experimental electronic structure data. For the band insulators and all other perovskites, a default value of 0.25 was used for the HF exchange fraction. For HSE calculations of doped perovskites, the fraction of exact exchange used was the same as the corresponding pure compound. While we do not know the precise optimal exact exchange value that should be used for each perovskite composition, select tests of the effect of the choice of the exact exchange fraction on the resulting work function for PrHfO$_3$ and Ba$_{0.5}$La$_{0.25}$Rb$_{0.25}$NbO$_3$ demonstrated that varying the exact exchange between 0.125 and 0.25 (the typical range of exact exchange fractions employed in HSE calculations) resulted in a change in work function on the order of 0.05 eV (see the **Supplementary Information** for more details).

Bulk perovskite structures were all simulated as stoichiometric $2 \times 2 \times 2$ supercells (40 atoms/cell). The use of $2 \times 2 \times 2$ supercells ensures that the octahedral tilting in the perovskite structure can be fully relaxed. When available, the bulk structures of undoped perovskites were modeled using initial structures contained in the Materials Project database.[65] The structures of doped perovskites were obtained by replacing elements in the pure perovskites at concentrations ranging from 12.5% to 50% of the A- and/or B-sites. Following the previous study of Jacobs *et al.*, to keep the number of screening calculations of doped compounds tractable, we considered a single ordering for each doped compound, where the dopant atoms were distributed as far from each other as possible on the appropriate perovskite sublattice (i.e. A- or B-sites). It is possible that additional concentrations could yield more interesting candidates. However, such studies become extremely computationally demanding as they would involve very large slabs with HSE. Therefore, we have not pursued additional concentrations at this time although they might be of interest for future work. We have confirmed that different orderings have only minor effects on the O *p*-band center and energy above convex hull (see the **Supplementary Information**). A total of 2913 bulk perovskite materials were examined in this work. Of these 2913 materials, 2514 materials were obtained from the work of Jacobs *et al.*, who used similar high-throughput DFT methods to search for stable, high activity perovskite oxides as solid oxide fuel cell cathodes.[39] The remaining 399 compounds were calculated in this study following preliminary screening analysis which demonstrated that metallic perovskites with a low number of *d* electrons tend to exhibit the lowest work functions, thus making this space of materials promising for additional, focused exploration. The source of each material is listed in the **Supplementary Information**.

The O *p*-band center is defined as the centroid of the electronic density of states projected onto the O 2*p* orbitals, referenced to Fermi level, and was calculated using Eq. (2):

$$\bar{O}_{2p}(E) = \frac{\int_{-\infty}^{\infty} E \cdot D_{O_{2p}}(E) dE}{\int_{-\infty}^{\infty} D_{O_{2p}}(E) dE} - E_{\text{Fermi}}, \qquad (2)$$

where $D_{O_{2p}}(E)$ stands for density of states projected onto the O 2*p* orbitals.[33]

In **Section 2.1.2** and **Section 2.3**, perovskite materials were analyzed by grouping materials based on the number of *d*-band electrons. The number of *d*-band electrons in a material was calculated using empirical electron counting rules and the assumption of completely ionic bonding. These assumptions result in alkali metals always regarded to be in the 1+ oxidation state, alkaline earth metals in the 2+ state, lanthanides in the 3+ state, oxygen in the 2- state, and the transition metals possessing an oxidation state or mix of oxidation states such that charge balance is realized. In the case of a material containing multiple species on the B-site, the number of *d*-band electrons for the material is taken as the average number of *d*-band electrons of the B-site species present. For example, in the case of SrVO$_3$, V is in the 4+ oxidation state, so this is a 3$d^1$ material, and the *d*-band occupation therefore equal to 1.

The work function ($\Phi$) is defined as the energy required to move an electron from the Fermi level ($E_{\text{Fermi}}$) to the vacuum level ($E_{\text{vacuum}}$):

$$\Phi = E_{\text{vac}} - E_{\text{Fermi}}. \tag{3}$$

$E_{\text{vacuum}}$ is defined as the value of the converged electrostatic potential sufficiently far from the terminating surface of interest such that the restoring force on an emitted electron is negligible. To obtain the work function for a given material, we constructed (001) orientated, AO-terminated symmetric 9-layer slabs (88 atoms per supercell). These slabs consist of 5 AO layers and 4 BO$_2$ layers, and have a 2 × 2 surface area containing 4 A or B cations per layer. The previous work of Jacobs *et al.* demonstrated that the use of 9-layer symmetric slabs was sufficient to give converged work functions with respect to slab thickness, and further, the work function values were not significantly different when symmetric versus asymmetric slabs of comparable thickness were used.[33] As the work function is sensitive to surface relaxation,[34] the atoms in the top and bottom 3 layers of the surface slabs were allowed to relax, and the remaining 3 layers of atoms were frozen to their bulk coordinates. Select tests at the GGA-level with 2 × 2 × 1 *k*-points showed that this choice of relaxing 3 layers resulted in changes of the calculated work function of less than 0.1 eV for SrNbO$_3$ (see the **Supplementary Information** for more details). All surface calculations were performed with dipole corrections enabled and at least a 20 Å vacuum region to minimize the interactions between surfaces. No surface reconstructions were explicitly added to the calculations, although relaxation was allowed to occur as described above. This approach will likely miss most surface reconstructions, especially those that require forming point defects, complex rumpling, or unit cells larger than used here. This limitation is significant, but more accurate treatment of reconstructions for large scale screening is not practical. We hope that the errors introduced by this approximation are at least somewhat ameliorated by our focus on conducting transition metal oxides, which can screen surface dipoles through charge rearrangement[68] without significant surface reconstruction. When surface reconstructions do occur, since we are most interested in behavior of the AO surface at high temperature and under high vacuum, common point defects are likely to be oxygen vacancies[69]–[71] or metal adatoms. Metal adatoms would almost certainly reduce the work function due to forming a positive dipole at the surface, thus making our calculated values upper bounds. Since our goal is low work function materials such an error would not change our list of promising materials. Oxygen vacancies seem to typically increase the work function but have a small effect at realistic concentrations[72], suggesting that the impact of such defects will not be significant and again not change our list of promising materials.

We note here that the use of computationally expensive HSE calculations appear to be required to obtain highly-accurate work functions for perovskite materials. Recent work from Ma *et al.*[72] showed through a detailed comparison of experimental and DFT-calculated work functions of an array of surface structures of $SrTiO_3$ that HSE and experimental work functions agree on average within about 0.2 eV. Further, it was found GGA can only capture relative differences in work functions between different types of surfaces (e.g. the difference in work function between AO- and $BO_2$-terminated surfaces) and does not provide the correct magnitude of the work function.[72]

To calculate the $E_{hull}$ values, we used the *GrandPotentialPhaseDiagram* class in the phase diagram module within the Pymatgen toolkit.[73] To calculate the $E_{hull}$, Pymatgen first establishes a phase diagram using all materials composed of given elements in the Materials Project database, then calculates the $E_{hull}$ at the point corresponding to the composition of the material under consideration. To mimic environmental conditions typical of a thermionic electron emission cathode, we calculated $E_{hull}$ for an open system where the chemical potential of O was calculated to coincide with thermionic cathode operating conditions of $T=1000$ °C and $p(O_2) = 10^{-10}$ Torr. The chemical potential of O was calculated following the work of Jacobs *et al.*, where the DFT-calculated $O_2$ molecule energy was shifted to account for the finite temperature gas enthalpy and entropy values at $T=1000$ °C, as well as the ideal gas partial pressure shift to coincide with $p(O_2) = 10^{-10}$ Torr.[74] Finally, the DFT-PBE errors associated with O binding were handled using the Materials Project energy shift of -0.7023 eV/O,[65] and the vibrational entropy contribution was included using an Einstein model with an Einstein temperature of 500 K, following previous studies.[12], [74]

**Supplementary Information:**

A spreadsheet is included as part of the Supplementary Information, which provides the materials comprising each material set during screening and their associated stability and O *p*-band center values. In addition, the Supplementary Information contains spreadsheets containing the data used to create each figure. Figures and associated data of the different convergence tests discussed in the Computational Methods section are also included in the Supplementary Information spreadsheet. Finally, the data mentioned above as well as the key VASP input and output files for all materials examined in this work are also publicly available on Figshare at http://doi.org/10.6084/m9.figshare.13020032.

**Conflicts of Interest:**

There are no conflicts to declare.

**Acknowledgements:**

This work was funded by the Defense Advanced Research Projects Agency (DARPA) through the Innovative Vacuum Electronic Science and Technology (INVEST) program. This work used the Extreme Science and Engineering Discovery Environment (XSEDE),[75] which is supported by National Science Foundation grant number ACI-1548562. This work used the XSEDE Stampede2 at the Texas Advanced Computing Center (TACC) through allocation TG-DMR090023. This research was performed using the compute resources and assistance of the UW-Madison Center for High Throughput Computing (CHTC) in the Department of Computer Sciences.